\documentclass[aps,pre,superscriptaddress,amsmath,amssymb,twocolumn]{revtex4}
\usepackage{graphicx}
\usepackage{hyperref}
\usepackage{color}
\definecolor{darkblue}{rgb}{0.0,0.0,0.5}
\hypersetup{colorlinks,breaklinks,
            linkcolor=darkblue,urlcolor=darkblue,
           anchorcolor=darkblue,citecolor=darkblue}

\usepackage[multiple]{footmisc}
\date{{}}

\allowdisplaybreaks

\begin{document}

\title{Suppression of Sub-surface Freezing in Free-Standing Thin Films of a Coarse-grained Model of Water}

\author{Amir Haji-Akbari}
%\email{hajakbar@princeton.edu}
\affiliation{Department of Chemical and Biological Engineering, Princeton University, Princeton, NJ 08544}

\author{Ryan S. DeFever}
\affiliation{Department of Chemical and Biomolecular Engineering, Clemson University, Clemson, SC 29634}

\author{Sapna Sarupria}
%\email{ ssarupr@g.clemson.edu}
\affiliation{Department of Chemical and Biomolecular Engineering, Clemson University, Clemson, SC 29634}

\author{Pablo G. Debenedetti}
\email{pdebene@exchange.princeton.edu}
\affiliation{Department of Chemical and Biological Engineering, Princeton University, Princeton, NJ 08544}

\date{\today}
\begin{abstract}
Freezing in the vicinity of water-vapor interfaces is of considerable interest to a wide range of disciplines, most notably the atmospheric sciences. In this work, we use molecular dynamics and two advanced sampling techniques, forward flux sampling and umbrella sampling, to study homogeneous nucleation of ice in free-standing thin films of supercooled water. We use a coarse-grained monoatomic model of water, known as mW, and  we find that in this model a vapor-liquid interface suppresses crystallization in its vicinity. This suppression occurs in the vicinity of flat interfaces where no net Laplace pressure in induced. Our free energy calculations reveal that the pre-critical crystalline nuclei that emerge near the interface are thermodynamically less stable than those that emerge in the bulk. We investigate the origin of this instability by computing the average asphericity of nuclei that form in different regions of the film, and observe that average asphericity increases closer to the interface, which is consistent with an increase in the free energy due to increased surface-to-volume ratios. 
\end{abstract}

\maketitle

\section{Introduction\label{section:intro}}
Water is arguably the most important molecule on earth. Its abundance in the biosphere, and its presence in the crystalline, liquid and gaseous states at conditions prevalent on Earth, is an important factor in the emergence and maintenance of life as we know it. In this context, the hydrologic cycle plays an indispensable role in promoting life~\cite{ThompsonGPC1989}, not only by maintaining biodiversity through the delivery of water throughout the earth, but also by sustaining a favorable climate without which most forms of life would cease to exist. It is therefore of utmost importance to understand the physical processes that constitute the hydrologic cycle.  One of the most important-- and probably the least understood-- is the formation of ice in the atmospheric droplets and aerosols that constitute clouds. The presence of icy droplets is not a pre-requisite for the formation of a cloud and in many climatological models, it is assumed that low-altitude and middle-altitude clouds are exclusively comprised of liquid droplets~\cite{ShineQJRMS1994}. However, the fraction and the distribution of frozen droplets in a cloud determines its overall properties. For instance, the radiative properties of icy and liquid droplets are significantly different. As a result, the fraction of frozen droplets in a cloud significantly affects its light-absorption properties, and is therefore an important factor in determining its radiation budget~\cite{ShineQJRMS1994,BakerScience1997}. Also, partially glaciated clouds are more likely to produce rainfalls than single-phase clouds made up of liquid droplets~\cite{KirkbyScience2002}. Due to these very important ramifications, the liquid fraction of a mixed-phase cloud is a very important input parameter to many climatological models~\cite{FowlerJClimate1996}. 

The problem of calculating the liquid fraction of mixed-phase clouds is however very challenging. Most existing models use empirical correlations to relate the ice content of a cloud to variables such as temperature~\cite{RaschJClimate1998}, while more sophisticated models use the liquid fraction as a prognostic variable that is directly computed from the model~\cite{WilsonQJRMS1999,RotstaynMWR2000}. However, all existing models perform poorly in predicting the correct liquid fraction of a cloud, and can sometimes underestimate it by a factor of two~\cite{BarrettJGR2013}.  This lack of predictive power arises from our lack of understanding of the molecular-level mechanisms that lead to ice formation in atmospheric droplets. From a thermodynamic perspective, ice formation is a \emph{first-order phase transition} and typically proceeds through a process known as \emph{nucleation and growth}. During nucleation, a so-called critical nucleus is formed in the supercooled liquid, such that smaller-sized nuclei dissolve spontaneously and larger-sized ice nuclei grow spontaneously. Subsequent growth of larger-than-critical nuclei is referred to simply as the growth process. In general nucleation is a fluctuation-driven rare event, and the probability of its occurrence decreases exponentially with the height of of the free energy barrier that separates the supercooled liquid and the crystalline basins. For pure water, these barriers can be relatively large, which makes the \emph{homogeneous nucleation} of ice very unlikely at temperatures close to the melting point. As a result, most of our day-to-day experiences of freezing occur through \emph{heterogeneous nucleation} in which an ice-nucleating particle facilitates freezing by decreasing the free energy barrier. It is indeed believed that ice formation in the atmosphere predominantly proceeds via heterogeneous nucleation mediated by impurities such as mineral dust, soot, biological, organic and ammonium sulfate particles~\cite{HooseAtmosChemPhys2012}. However, the amount of ice present in atmospheric clouds cannot be fully accounted for by heterogeneous nucleation alone~\cite{SassenJAtmosSci1988}. Therefore, both homogeneous and heterogeneous nucleation are important in determining cloud dynamics. On a molecular level, nucleation events-- whether homogeneous or heterogeneous-- generally occur at length ($\approx10^{-9}~$m) and time ($\approx10^{-9}~$s) scales that are not accessible to the existing experimental techniques, and there has only been success in measuring nucleation rates in narrow ranges of temperature without gaining any knowledge about the characteristics of the intermediate states~\cite{AndersonJAtmosSci1980, HagenJAtmosScie1981, MillerJCP1983, TaborekPRB1985, SassenJAtmosSci1988, HeymsfieldJAtmosSci1989, DeMottJAtmosSci1990, HeymsfieldJAtmosSci1993, KramerJCP1999, KoopNature2000,EarleAtmosChemPhys2010, KuhnAtmosChemPhys2011}.

One of the most important open questions in the area is whether a vapor-liquid interface facilitates or suppresses the formation of ice. This has been listed as one of the ten most important unknown questions about ice~\cite{RauschNature2013}. This controversy arises from the fundamental limitation of existing experimental techniques that are not yet capable of locating individual nuclei at their inception. Consequently, the evidence for the facilitation or suppression of crystallization are indirect. The idea of surface-facilitated crystallization was first proposed by Tabazadeh~\emph{et al}~\cite{TabazadehPNAS2002}. They used a simple thermodynamic reasoning to conclude that crystalline nuclei that form near the vapor-liquid interface will be thermodynamically favored over the nuclei emerging in the bulk if $\sigma_{sv}-\sigma_{lv}<\sigma_{ls}$, an inequality that they argue is satisfied for most single component systems. Here, $\sigma_{sv}, \sigma_{lv}$ and $\sigma_{ls}$ are the vapor-solid, vapor-liquid and solid-liquid surface tensions, and $\sigma_{sv}-\sigma_{lv}$ is the energetic penalty associated with forming a solid-vapor interface at the liquid-vapor interface. This inequality is equivalent to the condition that the liquid of a particular material wets its crystal partially, which is satisfied for most materials.  Using their  model, they re-analyzed some earlier experimental measurements of nucleation rates and were able to resolve apparent inconsistencies between those distinct measurements. However, they failed to back up their core thermodynamic argument with actual values for the liquid-vapor ($\sigma_{lv}$), solid-vapor ($\sigma_{sv}$) and solid-liquid ($\sigma_{sl}$) surface tensions of water, probably because of the difficulty in measuring these quantities at supercoolings relevant to atmospheric conditions. Further evidence for and against this theory emerged in later years, creating a controversy that is yet to be resolved~\cite{DuftACPD2004, ShawJPCB2005, GurganusJPhysChemLett2011}. For instance, Shaw~\emph{et al} observed several orders of magnitude increases in heterogeneous nucleation rates when the ice-nucleating particle was placed close to the vapor-liquid interface~\cite{ShawJPCB2005}.  However, Gurganus~\emph{et al} used optical microscopy to probe nucleation events in a water droplet placed on top of the surface of an ice-nucleating substrate, and observed no significant difference between the distribution of icy nuclei emerging at different regions of the surface~\cite{GurganusJPhysChemLett2011}. 
  Some authors have even suggested that the existing experimental techniques lack the necessary resolution for distinguishing surface- vs. volume-dominated nucleation~\cite{SignorellPhysRevE2008}.

In the absence of high-resolution experimental techniques, computer simulations are attractive alternatives for probing the length and time scales that are relevant in ice nucleation. However, computational studies of ice nucleation are also very challenging~\cite{BrukhnoJPhysCondMat2008}, and it was not until the turn of the millennium that Matsumoto~\emph{et al} were able to  nucleate ice in a molecular dynamics simulation of bulk supercooled water in the absence of any external stimuli-- such as electric fields-- or any biasing potentials~\cite{Matsumoto2002}. The microsecond-long trajectories that they obtained were the very first windows opened into the  molecular-level events that trigger ice nucleation. However, since there were only a handful of trajectories gathered in this study, it was not possible to explore the statistical nature of the nucleation process (e.g.~the most probable pathway of crystallization). For that, one  needs either to gather a large number of independent trajectories-- which is not usually practical--, or to use advanced molecular simulation techniques that sample the transition region of the configuration space in a targeted manner. Since then, numerous computational studies of ice nucleation have been performed, using a plethora of advanced sampling techniques and force fields~\cite{TroutJACS2003, QuigleyJCP2008, GalliPCCP2011, ReinhardtJCP2012, GalliNatComm2013, SanzJACS2013}. The simulation techniques used in many of these studies~\cite{TroutJACS2003, QuigleyJCP2008,  ReinhardtJCP2012} involve the application of a biasing potential. These techniques distort the true dynamics of the system, and are therefore not suitable for calculating kinetic properties such as nucleation rates. There is a second class of methods that sample the transition region without applying a biasing potential, and can thus be used for direct calculation of nucleation rates. In the context of ice nucleation, however, these methods have only been used for coarse-grained models of water. For instance, Li and coworkers~\cite{GalliPCCP2011, GalliNatComm2013} have computed homogeneous nucleation rates for the monoatomic water (mW) potential~\cite{MolineroJPCB2009}. 
However, applying these bias-free sampling techniques to molecular-- i.e.~multi-site-- models of water, such as the TIP4P family, remains an open challenge. Apart from large computational costs of estimating long-range electrostatic interactions, molecular models of water tend to have relaxation times that are orders of magnitude larger than  their coarse-grained counterparts. This latter fact makes structural relaxation of supercooled water always a source of concern in studies of ice nucleation. Indeed, the problem of calculating the rate of homogeneous ice nucleation for molecular models of water has been included among the most challenging problems in computational statistical physics, besting the efforts of large numbers of computational scientists~\cite{BrukhnoJPhysCondMat2008}.

Considering these challenges, it is not surprising that the problem of ice nucleation in the vicinity of vapor-liquid interfaces is yet to receive due scrutiny, and only a few computational studies have been performed~\cite{JungwithJPCB2006, JungwirthJPhysChemC2010, GalliNatComm2013}. With respect to the controversy of surface- vs.~volume-dominated nucleation, these studies reach  opposing conclusions. Jungwirth~\emph{et al.}~\cite{JungwithJPCB2006, JungwirthJPhysChemC2010} performed conventional MD simulations of free-standing thin films of a six-site model of water~\cite{NadaJCP2003}, and observed that nucleation events are more likely to occur in the vicinity of the vapor-liquid interface than the bulk. They explain this observation by arguing that electrostatic neutrality is violated in the vicinity of the vapor-liquid interface when the hydrogen atoms protrude towards the vapor phase. This, in turn, creates a net electric field in the interfacial region that enhances crystallization in the subsurface. Electrical fields are indeed know to enhance freezing~\cite{KusalikPRL1994}. In contrast, Li and coworkers~\cite{GalliNatComm2013} utilized the forward-flux sampling (FFS) algorithm~\cite{AllenFFSJCP2008} to calculate nucleation rates in nanodroplets of mW water~\cite{MolineroJPCB2009}, and they observed a dramatic decrease in nucleation rates compared to the bulk. This observation was attributed to the presence of a large Laplace pressure induced inside those droplets that leads to a decrease in nucleation rates in materials that have negative-slope melting curves. 

In this work, we use a range of molecular simulation techniques to study homogeneous nucleation of ice in free-standing thin films of supercooled water. We first carry out multiple conventional molecular dynamics simulations of films of mW water at 200~K and observe that freezing events are more likely to start in the bulk than in the subsurface region.  We then use the forward flux sampling algorithm to explicitly calculate nucleation rates both in the bulk and in the free-standing thin films at temperatures between 220 and 235~K, and observe a two- to three orders of magnitude decrease in nucleation rates in 5 nm-thick films. We then compute the reversible work of formation for crystalline nuclei of different sizes as a function of distance from the vapor-liquid interface, and observe that the clusters in the bulk are favored over the clusters that are close to the surface. Finally, we elaborate on the origin of the suppression of crystallization in the vicinity of the vapor-liquid interface by analyzing the geometric shapes of crystalline clusters and by investigating the structural and dynamical features of the interface. 

\section{Methods}

\subsection{Water Model}
We represent water molecules using the mW potential~\cite{MolineroJPCB2009}, which is based on the Stillinger-Weber force field, originally developed for simulating Group IV elements such as carbon and silicon~\cite{StillingerPRB1985}. The mW potential preserves the Stillinger-Weber form, but has been parametrized to reproduce thermodynamic and structural properties of water~\cite{MolineroJPCB2009}. An mW water molecule has no hydrogens or oxygens, and as a result, no long-range electrostatic interactions need to be computed during the simulation. Instead, the existence of the hydrogen bond network is implicitly mimicked by including  a three-body term that favors locally tetrahedral arrangements of water molecules. Due to the lack of electrostatic interactions, this model accelerates water dynamics (e.g.~it overestimates the self-diffusion coefficient~\cite{MolineroJPCB2009}) even though it successfully predicts the structure, the energetics, and the anomalies of water. It is because of this speeding up of dynamics that the rate of homogenous ice nucleation can be readily computed for the mW system~\cite{GalliPCCP2011}, unlike most molecular models of water for which no explicit direct rate calculations have been reported. Despite the 'fast` dynamics of the mW model, the key assumption underlying this work is that such overestimations will essentially cancel out when comparing the nucleation rates in films and in the bulk. In other words, we are interested in comparing bulk and surface nucleation rates rather than predicting absolute nucleation rates that are relevant to real water.

\subsection{System Preparation and Molecular Dynamics Simulations}
We carry out our simulations in cuboidal boxes that are periodic in all dimensions. For ice nucleation in the bulk, we use cubic boxes that contain $2^{12}=4096$ water molecules. The starting configurations are prepared by constructing a dilute simple cubic lattice of mW molecules, followed by rapidly compressing it to the target temperature and pressure with a nanosecond-long molecular dynamics simulation in the $NpT$ ensemble. For ice nucleation in free-standing thin films, the cuboidal boxes are stretched along the $z$ direction, and the initial configurations are obtained by taking the configurations prepared for the bulk simulations, and expanding the simulation box in the $z$ direction by a factor of five. This is to assure that the films are not affected by their periodic images. The arising configurations are then equilibrated in a nanosecond-long MD simulation in the $NVT$ ensemble.

We perform our molecular dynamics simulations using LAMMPS~\cite{PlimptonJCompPhys1995}. Newton's equations of motion are integrated using the velocity Verlet algorithm~\cite{SwopeJCP1982} with a time step of $\Delta{t}=2$~fs, and temperature and pressure are controlled using a Nos\'{e}-Hoover thermostat ($\tau=0.2~\text{ps}$)~\cite{NoseMolPhys1984,HooverPhysRevA1985} and a Parrinello-Rahman barostat ($\tau=2.0~\text{ps}$)~\cite{ParrinelloJAppPhys1981} respectively. 

\subsection{Order Parameter}
A crucial component of any computational investigation of crystallization is the order parameter that is used for quantifying the progress of crystallization. For this purpose, two classes of order parameters are used that are both based on the bond orientational order parameters of Nelson and Toner~\cite{Nelson1981PhysRev_BoundOP}. The procedure starts by identifying the neighbors of every molecule in the system, based on a distance criterion. Then, spherical harmonics are used for quantifying the relative arrangement of neighbors of every given molecule by computing:
\begin{eqnarray}
	q_{lm}(i) &=& \frac{1}{N_b(i)}\sum_{j=1}^{N_b(i)}Y_{lm}(\theta_{ij},\phi_{ij})\label{eq:qlm}
\end{eqnarray}
where $N_b(i)$ is the number of neighbors of the $i$th particle, $\theta_{ij}$ and $\phi_{ij}$ are the spherical angles associated with the displacement vector connecting the $i$th particle to its $j$th neighbor, and $Y_{lm}(\cdot,\cdot)$ is the spherical harmonic given by:
\begin{eqnarray}
Y_{lm}(\theta,\phi)&=& \sqrt{\frac{2l+1}{4\pi}\frac{(l-m)!}{(l+m)!}}P_l^m(\cos\theta)e^{im\phi}
\end{eqnarray}
with $l=0,1,2,\cdots$ and $m=-l,-l+1,\cdots,l-1,l$, and $P_l^m(\cdot)$ the associate Legendre polynomial. Based on the type of order present in the system, one or two values of $l$ are used. The two classes of order parameters differ on how the individual $q_{lm}$ values are combined to quantify the long-range translational order in the system. In the first class of order parameters known as \emph{global order parameters}, individual $q_{lm}(i)$'s are averaged to form a set of global $Q_{lm}$'s that are then used for computing scalar invariants that quantify the extent of crystallization in the system. On the contrary, \emph{local order parameters} are based on identifying the types (solid-like vs.~liquid-like) of individual   molecules by computing those scalar invariants for every individual molecule. A graph of neighboring solid-like molecules is then constructed in the system to form \emph{clusters} of solid-like molecules. In studies of ice nucleation, global order parameters have been historically used when a biasing potential is applied for constructing a reversible thermodynamic path that connects the crystalline and the amorphous basins~\cite{TroutJACS2003}, while local order parameters are typically used in situations when no biasing potential is employed~\cite{Matsumoto2002}. 

In this work, we use the local $q_6$ order parameter as explained in Ref.~\cite{GalliPCCP2011}. A nearest neighbor shell of $3.2$~\AA~in radius is used for identifying the neighbors of each molecule. The $q_{6m}$'s are then calculated for each molecule using Eq.~(\ref{eq:qlm}), and the local $q_6$ order parameter is calculated as:
\begin{eqnarray}
q_6(i) &=& \frac{1}{N_b(i)}\sum_{j=1}^{N_b(i)}\frac{\textbf{q}_6(i)\cdot\textbf{q}^*_6(j)}{|\textbf{q}_6(i)|\cdot|\textbf{q}_6(j)|}
\end{eqnarray}
Here, $\textbf{q}_6(i)$ is a vector that contains all thirteen $q_{6m}$ elements, and $\textbf{a}\cdot\textbf{b}^*$ is the inner product of vector $\textbf{a}$ and the complex conjugate of vector $\textbf{b}$. The $i$th molecule is classified as solid-like if $q_6(i)>0.5$~\cite{GalliPCCP2011}. In order to remove chains of locally tetrahedral water molecules that are widely present in supercooled water (as opposed to compact arrangements that are physically relevant to the ice nucleation process), the chain exclusion algorithm of Reinhardt~\emph{et al}~\cite{VegaJCP2012} is used to further refine the identity of solid-like molecules, as follows. First, every solid-like molecule that has more than four nearest neighbors is labeled as 'liquid-like`. Then, a graph is constructed by recursively connecting the remaining solid-like molecules to their solid-like neighbors. The arising graph is further refined by excluding the solid-like molecules that have one solid-like neighbor only unless that one solid-like neighbor is connected to a minimum of three solid-like molecules. This latter step is only performed on clusters that have a minimum of ten water molecules. The size of the largest surviving cluster of solid-like molecules $\lambda$ is used as the order parameter to quantify the progress of crystallization. Throughout this work, we will also refer to this largest cluster of solid-like molecules as the largest crystalline nucleus.

\subsection{Forward-Flux Sampling}
Among the advanced sampling techniques that can be used for direct calculation of nucleation rates~\cite{DallagoChandlerJCP1998, ChandlerTPS2002, BolhuisJCompPhys2005, AllenFrenkel2006}, forward-flux sampling (FFS)~\cite{AllenFrenkel2006} is the least sensitive to the proper selection of the order parameter. This is a considerable advantage in studying a process as complicated as crystallization for which the \emph{a priori} identification of a good order parameter is not trivial.
Not surprisingly, forward-flux sampling has gained  popularity in recent years, and has been successfully used for computing crystallization rates in systems such as hard spheres~\cite{FillionJCP2010}, silicon~\cite{LiNatMater2009}, NaCl~\cite{ValerianiJCP2005}, oppositely-charged colloidal particles~\cite{SanzPRL2007} and coarse-grained water~\cite{GalliPCCP2011,GalliNatComm2013}. The basic idea of the FFS algorithm is to partition the configuration space into non-overlapping regions that are divided by the isosurfaces of the order parameter referred to as \emph{milestones}.  
The closest milestone to the liquid basin, denoted by $\lambda_{\text{basin}}$, is chosen so that it is frequently crossed by the configurations sampled from the supercooled liquid basin. The other milestones are chosen so that every one of them is accessible frequently enough to the trajectories that are initiated at the previous milestone. The nucleation rate is then expressed as:
\begin{eqnarray}
R &=& \Phi_0\prod_{i=1}^N\mathbb{P}(\lambda_i|\lambda_{i-1})
\end{eqnarray} 
where $\Phi_0$ is the flux of trajectories that cross the zeroth milestone, and $\mathbb{P}(\lambda_i|\lambda_{i-1})$ is the probability that a trajectory that is initiated from a configuration at the $(i-1)$th milestone crosses the $i$th milestone \emph{before} returning to the liquid basin. An FFS calculation is terminated when $\mathbb{P}(\lambda_N|\lambda_{N-1})\equiv1$ for every $\lambda_N>\lambda_{N-1}$. This means that the configurations gathered at $\lambda_{N-1}$ are all post-critical and therefore  always grow with probability one irrespective to the position of the next milestone. In order to compute the flux, a series of long MD simulations are carried out in the basin and the configuration of the system is stored whenever the zeroth milestone is crossed. Those configurations are then used in the second stage of the algorithm to calculate $\mathbb{P}(\lambda_1|\lambda_0)$ in a Monte Carlo scheme carried out as follows: A configuration is randomly chosen from among the configurations at $\lambda_0$. The momenta of the molecules are randomized according to the Boltzmann distribution, and the system is evolved using Hamiltonian dynamics.  The arising MD trajectory is terminated either if it crosses $\lambda_1$ or if it returns back to the liquid basin. The configurations of the system in successful crossings of $\lambda_1$ are stored for future iterations, and $\mathbb{P}(\lambda_1|\lambda_0)$ is computed as the fraction of trajectories that cross $\lambda_1$ before returning to the liquid basin. The same procedure is repeated for the configurations gathered at $\lambda_1,\lambda_2,\cdots$, until a value of $\lambda_N$ for which $\mathbb{P}(\lambda_{N+1}|\lambda_{N})$ converges to unity. For every $\lambda\in\{\lambda_1,\cdots,\lambda_N\}$, the cumulative transition probability is defined as $\mathbb{P}(\lambda|\lambda_0):=\prod_{k=1}^{i}\mathbb{P}(\lambda_{k}|\lambda_{k-1})$.

We carry out all the stages of our FFS rate calculations using an in-house C++ program. This program links against the LAMMPS static library and employs it as its internal MD engine. For rate calculations in the bulk, the individual MD trajectories are carried out in the $NpT$ ensemble at $p=1$~bar, while rate calculations in the films are performed with trajectories in the $NVT$ ensemble. For every rate calculation, we choose $\lambda_{\text{basin}}$ and $\lambda_0$ as follows. If $\psi(\lambda)$ is the equilibrium distribution of the order parameter in the supercooled liquid basin with mean $\mu$ and standard deviation $\sigma$, we choose $\lambda_{\text{basin}}$ to be an integer between $\mu$ and $\mu+\sigma$. A suitable value of $\lambda_0$ is chosen so that $10^{-3}\le\sum_{n=\lambda_0}^{\infty}\psi(n)\le10^{-2}$. The flux is then calculated as $\Phi_0=N_{\text{cross}}/t\langle{V}\rangle$ with $N_{\text{cross}}$ the number of successful crossings, $t$ the length of the MD trajectory, and $\langle V\rangle$  the average volume of the liquid region. A crossing is defined as successful if $\lambda_0$ is crossed by a trajectory originating from $\lambda_{\text{basin}}$. In the case of rate calculations in the bulk, $\langle V\rangle$ is the average volume of the system, while for free-standing thin films, $\langle V\rangle$ is computed by partitioning the simulation box into a grid of cubic cells of side $3.2~\AA$, and by  enumerating the average number of cells that have at least eleven non-empty neighboring cells. 

After computing $\Phi_0$ and gathering a sufficient number of configurations at $\lambda_0$, we use those for computing transition probabilities. The exact locations of the remaining milestones are determined so that for every two consecutive milestones, the transition probability is between $10^{-3}$ and $10^{-1}$, except for the very last two milestones in which transition probabilities are $\ge1/2$. We terminate each iteration after observing a minimum of $700$ successful crossings.
We request more crossings if the transition probability is smaller in order to decrease the relative statistical error in the estimate of the corresponding $\mathbb{P}(\lambda_i|\lambda_{i-1})$.

\subsection{Umbrella Sampling}
In order to compute the free energy of formation for clusters of different sizes as a function of distance from the surface, we consider a 5-nm-thick film of $4\,096$ water molecules at $220$~K, and perform the umbrella sampling simulations~\cite{TorrieJCompPhys1977} using the following biasing potential:
\begin{eqnarray}
U_{i,\text{bias}}(\textbf{r}^N) &=& \tfrac12k_{{\lambda},i}\big[\lambda(\textbf{r}^N)-\lambda_i\big]^2+\tfrac12k_{z,i}\big[ z(\textbf{r}^N)-z_i\big]^2\notag\\
&&
\end{eqnarray}
where $\lambda(\textbf{r}^N)$ is the size of the largest solid-like cluster in the system, and $z(\textbf{r}^N)$ is the distance of the center of mass of the largest  cluster from the center of the film. $\lambda_i$ and $z_i$ are the target values of $\lambda$ and $z$ in the $i$th umbrella sampling simulation. We perform these calculations at 220~K since the nucleation barrier is expected to be smaller at 220~K than the other temperatures at which rate calculations are performed. We carry out a total of $350$ distinct umbrella sampling simulations spanning the range of $0\le z\le 24~\AA$ and $0\le\lambda\le284$, and combine the resulting histograms using the weighted histogram analysis method (WHAM)~\cite{KumarWHAM1992, KumarWHAM2004}. Due to the discontinuous nature of the order parameter, it is not possible to sample the biased energy landscape using molecular dynamics. Instead, we use a hybrid Monte Carlo scheme~\cite{DuanePhysLettB1987} in which short $NVE$ MD trajectories act as trial moves of the Monte Carlo simulation, with the move being accepted or rejected according to the Metropolis criterion. Each such MD trajectory is comprised of two MD steps, with step sizes ranging between $2$ and $30$~fs. The step size is occasionally adjusted during the simulation in order to achieve a target acceptance probability of $0.4$.  

It is necessary to mention that we do not start our umbrella sampling simulations from configurations that are obtained from the forward-flux sampling. Instead, we
 initiate our umbrella sampling simulations at low values of $\lambda$-- i.e.~$\lambda=5$-- by taking suitable configurations from our basin simulations. All other umbrella sampling simulations use a starting configuration that has been generated in the umbrella sampling simulation conducted at a neighboring window, i.e.~with equal $\lambda$ and different $z$ value, or different $\lambda$ and an equal $z$ value.

\section{Results and Discussions\label{section:results}}

\begin{figure}
\begin{center}
	\includegraphics[width=.5\textwidth]{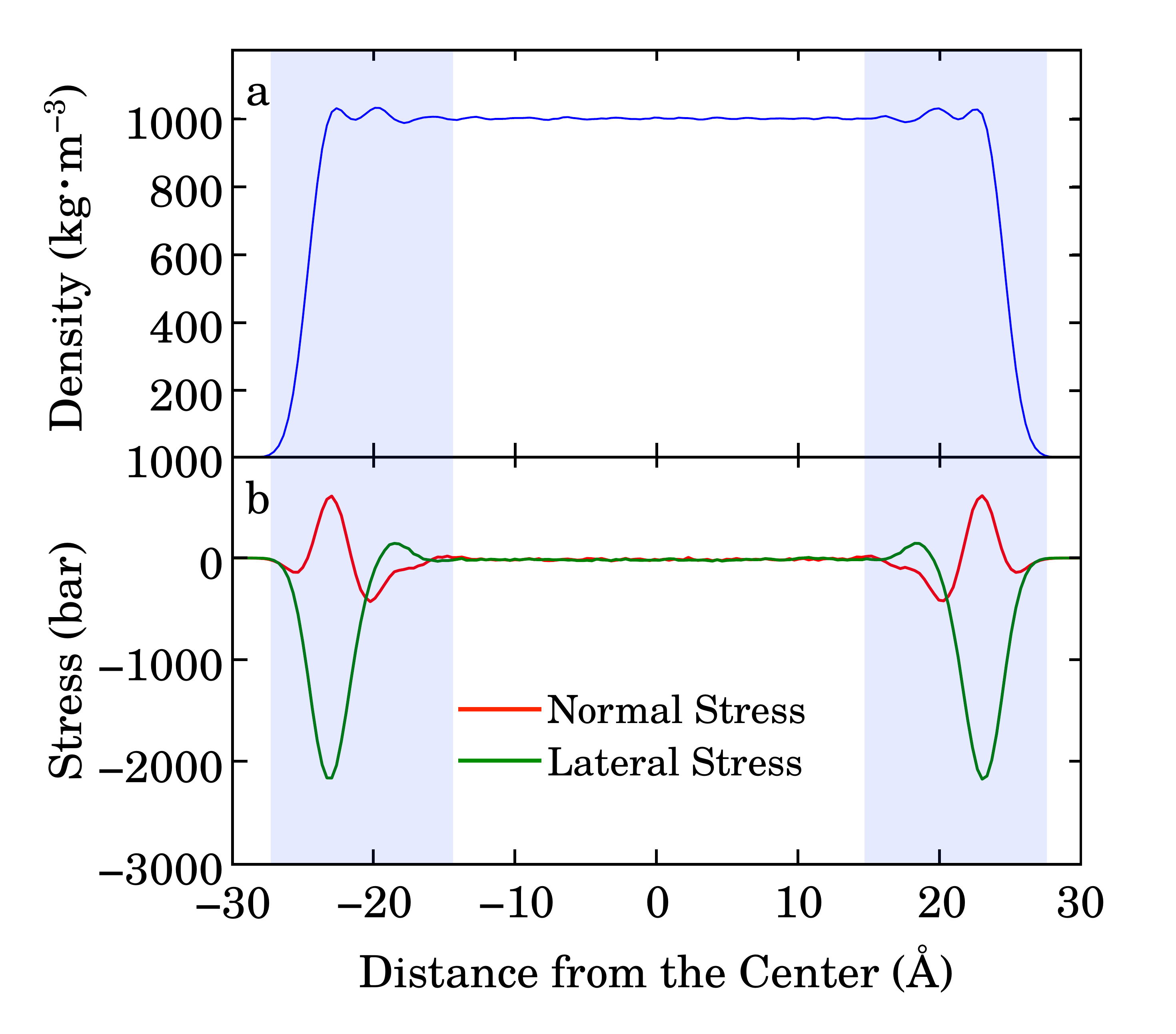}
	\caption{(a) Density and (b) stress profiles across the 5-nm film at $220$~K.\label{fig:densitystress}}
\end{center}
\end{figure}

\subsection{Identification of the Subsurface Region}

Before studying crystallization in free-standing thin films of supercooled water, we first need to identify a suitable definition for the subsurface region, or the region of a film that is affected by the presence of the vapor-liquid interface. We do this by computing the profiles of several thermodynamic and kinetic properties, such as density, stress and relaxation time, across the film using molecular dynamics simulations.  These calculations are performed using another in-house computer program of ours described elsewhere~\cite{HajiAkbariJCP2014}. Fig.~\ref{fig:densitystress} depicts  profiles of density and lateral and normal stress for a liquid film at $220$~K. The deviations of density and stress from the bulk values are only significant in a region that is around 12~\AA~thick. In Fig.~\ref{fig:densitystress}, this region is depicted in shaded blue. The same behavior is observed in the films simulated at other temperatures. We therefore define the subsurface region as a buffer zone that is 12~\AA~in thickness, for all the films studied in this work.

\begin{figure*}
\begin{center}
	\includegraphics[width=\textwidth]{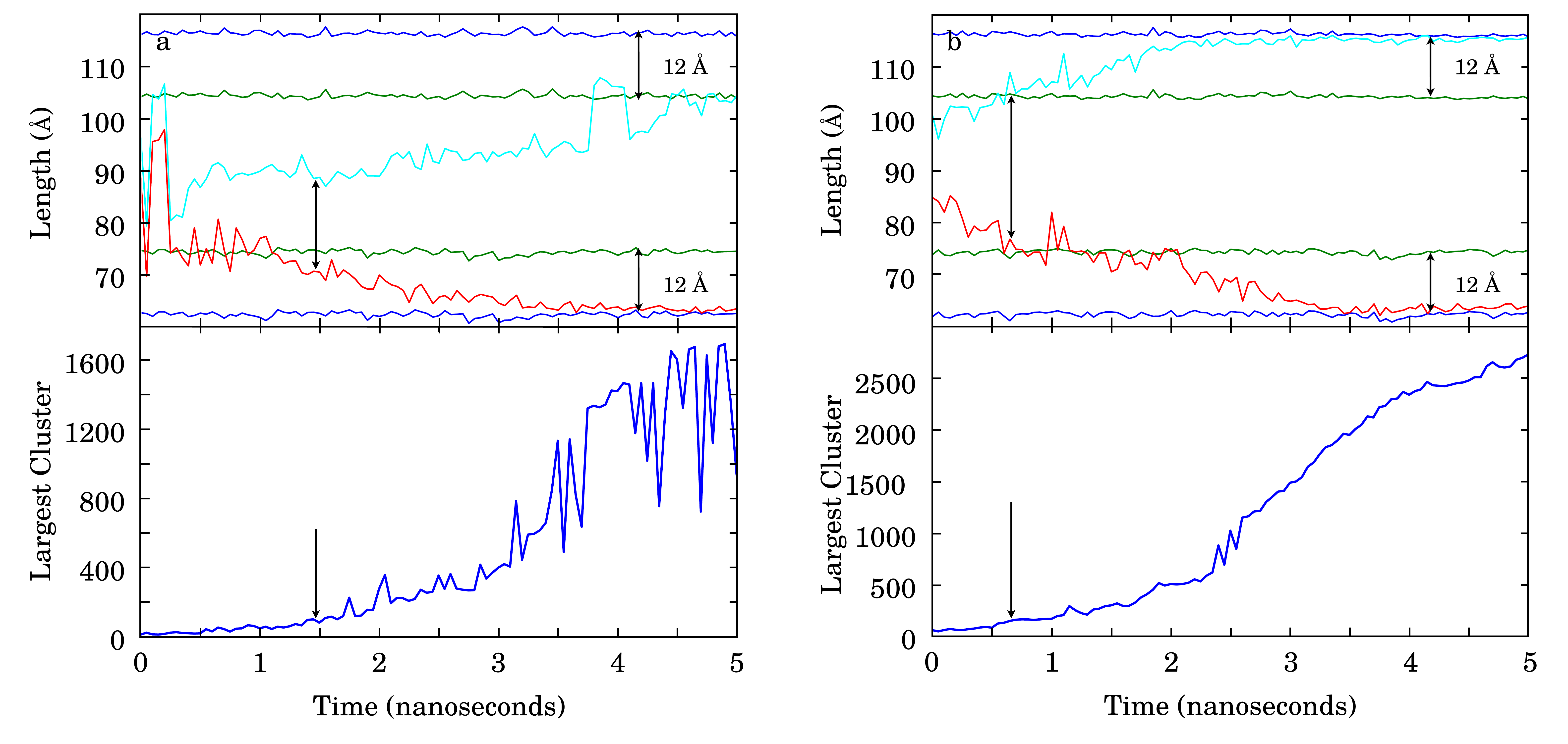}
	\caption{Examples of (a) surface, and (b) bulk crystallization in conventional MD simulations of mW films at $200$~K. The top panels show the geometric boundaries of the film (blue), the subsurface region (green) and the largest solid-like cluster (red and cyan). The bottom panels show the size of the largest solid-like cluster. In panel a, the black arrow corresponds to the time after which fluctuations in the cluster size are characterized by the sudden accretion or loss of large numbers of particles (peaks), superimposed on an overall accelerating growth.. In panel b, the black arrow corresponds to the time at which the largest cluster penetrates into the subsurface region for the first time. As explained in the text, by this time the cluster is post-critical. \label{fig:bruteMDplots}}
\end{center}
\end{figure*}

\subsection{Conventional MD Simulations at 200~K}
\begin{figure}
	\begin{center}
		\includegraphics[width=.45\textwidth]{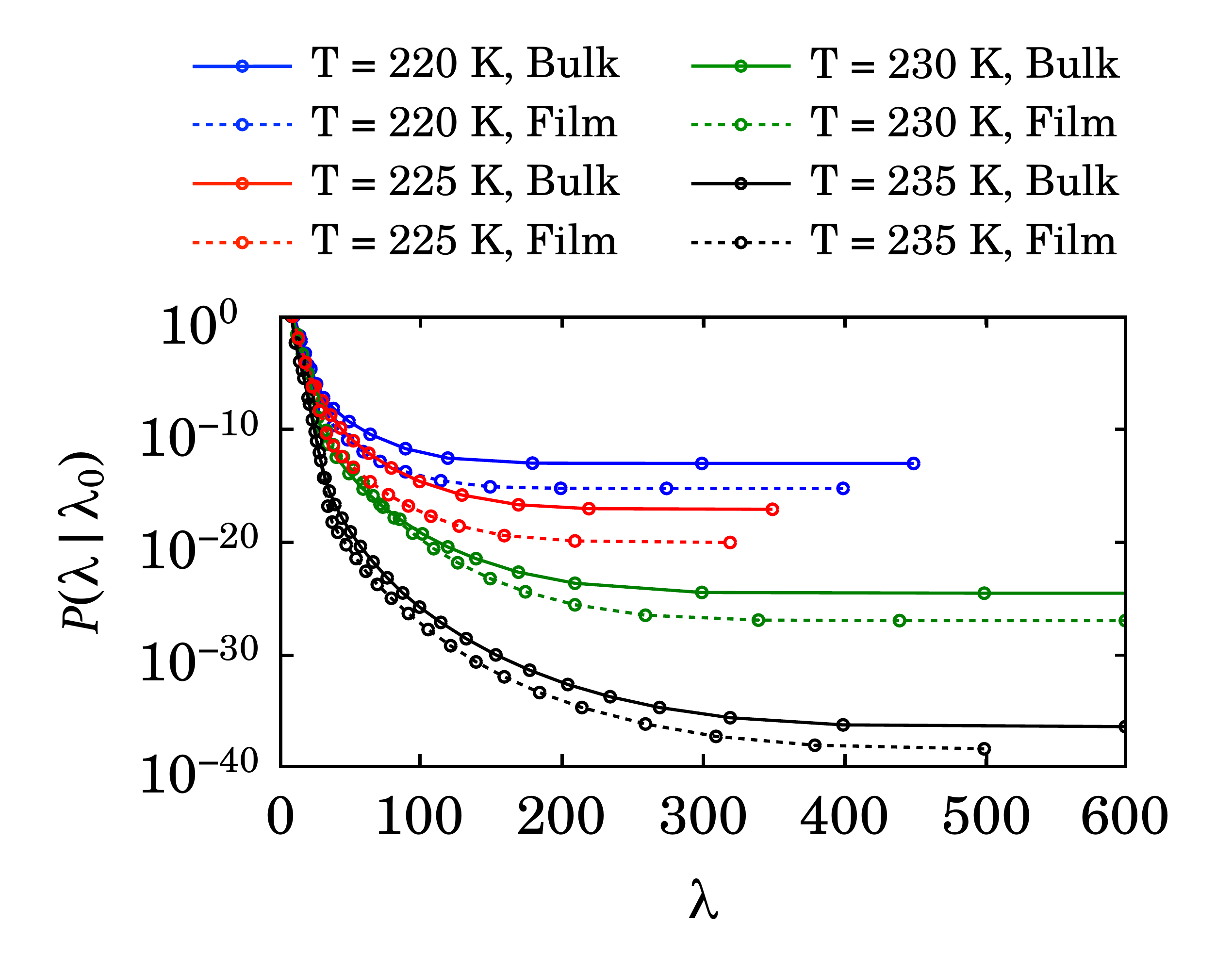}
		\caption{\label{fig:prob} Cumulative transition probability, $\mathbb{P}(\lambda|\lambda_0)$ vs.~$\lambda$ for FFS calculations of the nucleation rate in the bulk supercooled mW water, as well in films that are 5-nm thick.}
	\end{center}
\end{figure}

After obtaining a reasonable definition of the subsurface region, we carry out conventional MD simulations of liquid mW films at $200$~K, the temperature at which ice nucleation is the fastest for the mW potential~\cite{MolineroNature2011}. We then enumerate the number of crystallization events that start in the bulk vs.~the ones that start in the subsurface region. In order to do that, we take $49$ independent configurations for our mW film simulated at $220$~K, and gradually quench them down to $200$~K in eight-nanosecond-long $NVT$ MD simulations. We then equilibrate those configurations at $200$~K for $92$ additional nanoseconds, and monitor the crystallization by analyzing the configurations gathered  every $50~\text{ps}$. For each configuration, we compute the size of the largest cluster as well as its geometric boundaries as defined by the minimum and the maximum $z$ coordinate of the molecules in the cluster. Fig.~\ref{fig:bruteMDplots} depicts two such trajectories that crystallize within the first five nanoseconds of the equilibration simulations. In Fig.~\ref{fig:bruteMDplots}a, the growing crystalline nucleus resides partly in the subsurface region of the film. The black arrow marks the approximate time at which fluctuations in the size of the cluster become significantly enhanced, and the growth process accordingly becomes characterized by the rapid accretion or loss of large numbers of particles (peaks), superimposed on the overall accelerated size increase. At that point, the crystalline nucleus partially resides in the subsurface region.  In Fig.~\ref{fig:bruteMDplots}b, however, the nucleus completely resides in the bulk region of the film. Indeed the moment the largest cluster penetrates into the subsurface region for the first time (the black arrow at Fig.~\ref{fig:bruteMDplots}b), it is comprised of around 200 molecules. This is  close to the critical cluster size at 220~K (see Fig.~\ref{fig:prob}), so one would expect that such a cluster will be post critical at 200~K. (Refer to the discussion of Fig.~\ref{fig:prob} in Section~\ref{section:results:ffs} for further discussion on how critical nucleus sizes are determined from FFS calculations.)  This clearly shows that nucleation has started completely in the bulk for this trajectory.
 We classify the first trajectory as an example of  'surface` crystallization while the second trajectory  is counted as an instance of 'bulk` crystallization. From the 49 trajectories studied, four of them crystallized during the initial quenching period. From the remaining 45 trajectories, crystallization started in the subsurface region in only 13 of them. This observation is an indication that vapor-liquid interfaces suppress crystallization in the mW system. 

\begin{figure}
	\begin{center}
		\includegraphics[width=.5\textwidth]{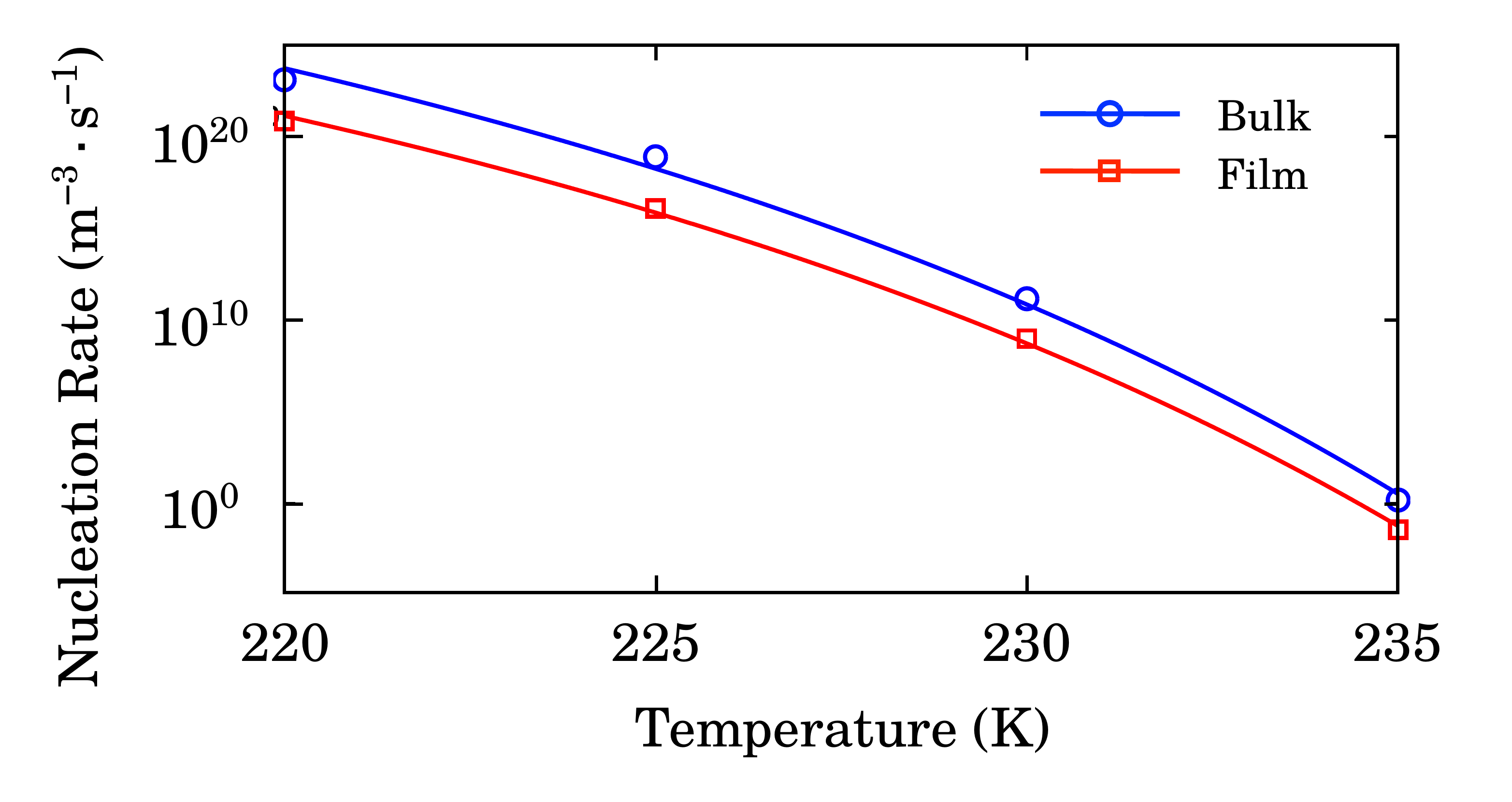}
		\caption{Temperature dependence of computed nucleation rates.\label{fig:rates}}
	\end{center}
\end{figure}

\begin{figure}
	\begin{center}
		\includegraphics[width=.5\textwidth]{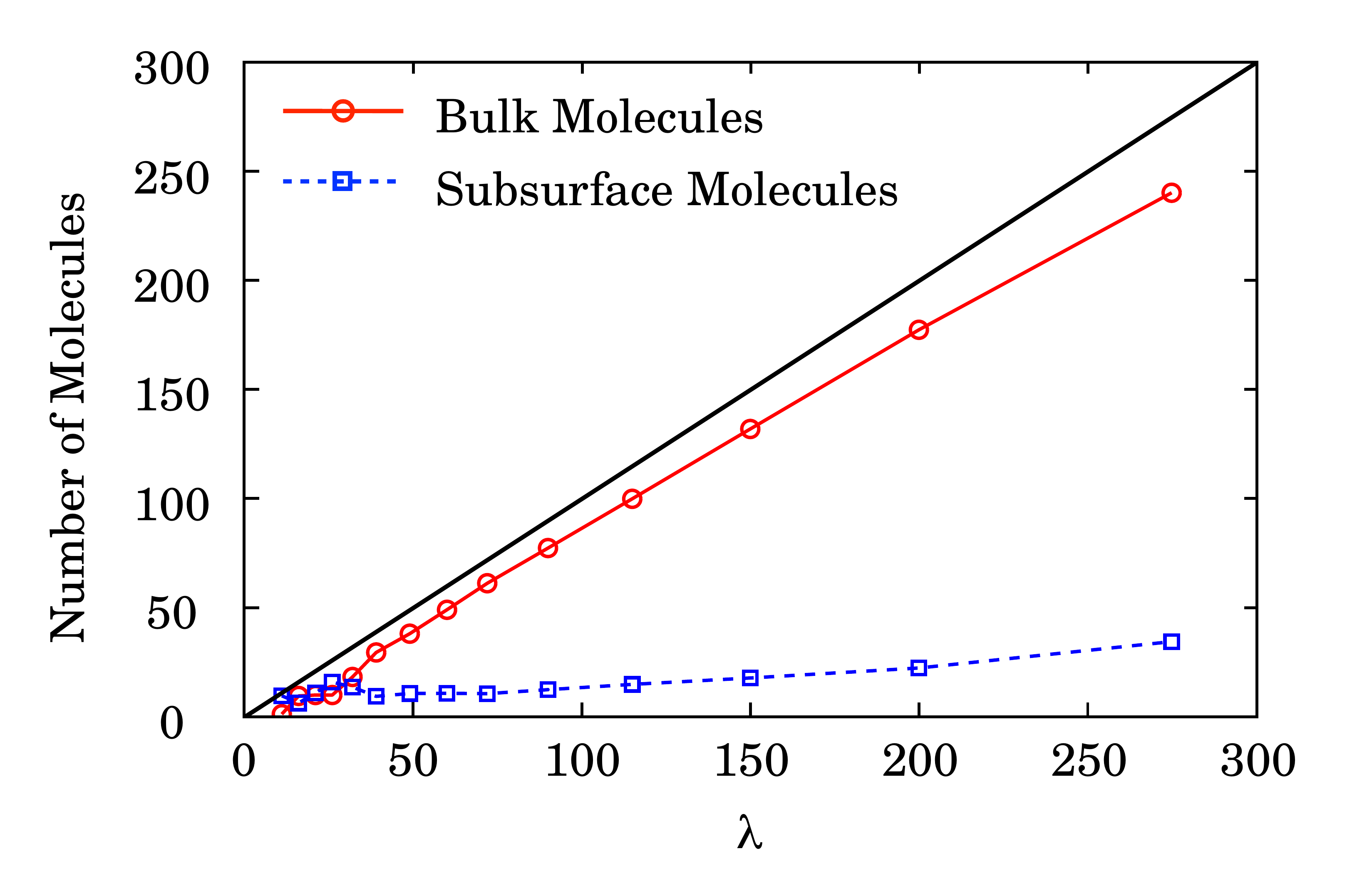}
		\caption{Average number of molecules belonging to crystalline nuclei of size $\lambda$ that reside in the bulk region (solid red) and in the subsurface region (dashed blue). The analysis is performed for the configurations gathered during the FFS calculations of nucleation rate for the 5-nm film at 220~K. The solid dark line has a slope of unity. In every configuration, around $2\,000$ water molecules are located in the subsurface region. Each datapoint has been obtained from a minimum of 500 snapshots, and the error bars are all smaller than the size of the symbols.  \label{fig:bulkfilm220}
		}
	\end{center}
\end{figure}

\begin{figure}
	\begin{center}
		\includegraphics[width=.45\textwidth]{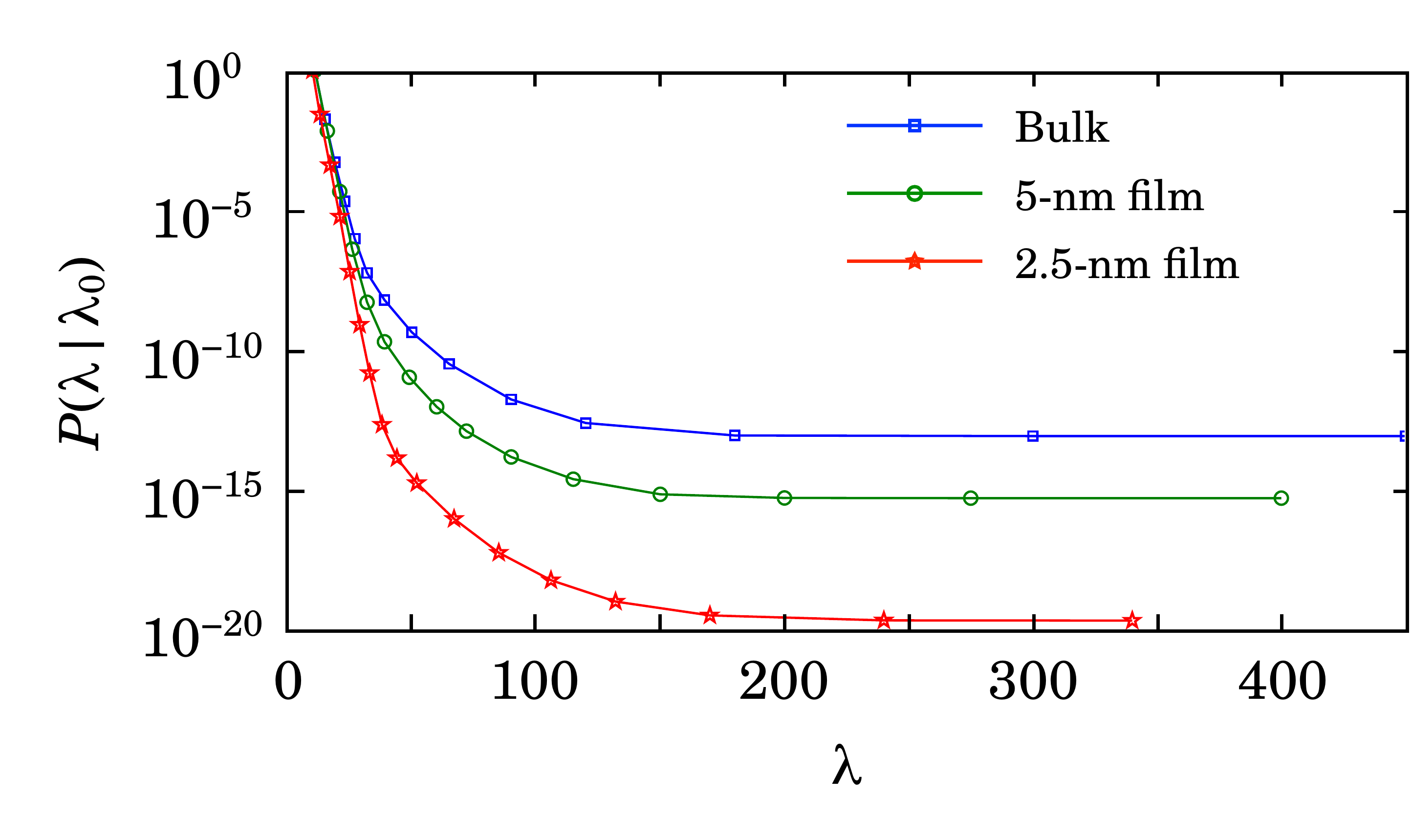}
		\caption{\label{fig:thickness}Cumulative transition probability, $\mathbb{P}(\lambda|\lambda_0)$ vs.~$\lambda$ for FFS calculations in films of different thicknesses at 220~K.}
	\end{center}
\end{figure}

\begin{table*}
	\centering
	\begin{minipage}{5.6in}
	\caption{\label{tab:flux}Computed fluxes in FFS calculations of nucleation rates. }
	\begin{tabular}{lcccrcccc}
		\hline\hline
		System~ & ~$T~(\text{K})$~ & ~~$\lambda_{\text{basin}}$~~ & ~~$\lambda_0$~~ & ~ $t~(\text{ns})$~ & ~$N_{\text{cross}}$~ & ~$\langle V\rangle~(\text{nm}^{-3})$~ &  ~$\Phi_0~(\text{m}^{-3}\cdot\text{s}^{-1})$~ & ~$\epsilon_{\log_{10}\Phi_0}$\footnote{$\epsilon_{\log_{10}\Phi_0}$ is the absolute error in the decimal logarithm of $\Phi_0$.}~ \\
		\hline 
		Film, 5 nm & 220 & 6 & 11 & 39.800 & $6\,517$ & 125.177 & $1.307\times10^{36}$ & $0.0108$ \\
		Film, 5~nm & 225 & 6 & 10 & 75.070 & $12\,602$ & $124.771$ & $1.346\times10^{36}$ & $0.0080$   \\
		Film, 5~nm & 230 & 5 & 10 & 34.048 & $3\,709$ & $124.811$ & $8.728\times10^{35}$ & $0.0144$ \\
		Film, 5~nm & 235 & 5 & 9 & 34.998 & $31\,600$ & $124.922$ & $7.228\times10^{36}$ & 0.0048\\
		Film, 2.5-nm & 220 & 5 & 10 & 269.276 & $41\,500$ & $~63.841$ & $2.414\times10^{36}$ & 0.0042 \\
		Bulk & 220 & 6 & 11 & $69.070$ & $12\,194$ & $122.894$ & $1.437\times10^{36}$ & $0.0080$\\
		Bulk & 225 & 6 & 10 & $102.623$ & $13\,015$ & $122.598$ & $1.034\times10^{36}$ & $0.0076$\\
		Bulk & 230 & 5 & 10 & $67.730$ & $4\,109$ & $122.397$ & $4.957\times10^{35}$ & $0.0136$ \\
		Bulk & 235 & 4 & 9 & $121.508$ & $48\,294$ & $122.260$ & $3.251\times10^{36}$ & $0.0040$\\
		\hline
	\end{tabular}
	\end{minipage}
\end{table*}

\begin{table*}
	\centering
	\begin{minipage}{5.8in}
	\caption{\label{tab:rates}Transition probabilities and nucleation rates for the systems considered in this work.}
	\begin{tabular}{lcccr}
		\hline\hline
		System~~~~ & ~~$T~(\text{K})$~~ & ~~~~~$\log_{10}\Phi_0$~~~~~ & ~~~~~$\log_{10}\mathbb{P}(\lambda_{x}|\lambda_0)\footnote{$\lambda_x$ corresponds to the value of the order parameter that completely lies in the crystalline basin.}\textsuperscript{,}\footnote{Statistical uncertainties in transition probabilities are computed using the procedure described in Ref.~\cite{AllenFFSJCP2006} and correspond to 95~\% confidence intervals.}$~~~~~ & ~~~~~$\log_{10}R$\footnote{Like $\Phi_0$, $R$ has the units of $\text{m}^{-3}\cdot\text{s}^{-1}$.}~~~~~~   \\
		\hline 
		Film, 5 nm & 220 &  $36.1163\pm0.0108$ & $-15.2638\pm0.3498$ & $20.8525\pm0.3498$ \\
		Film, 5~nm & 225 &  $36.1290\pm0.0080$ & $-20.0583\pm0.4426$ & $16.0707\pm0.4426$  \\
		Film, 5~nm & 230 & $35.9409\pm0.0144$  & $-26.9815\pm0.4690$ & $8.9594\pm0.4690$\\
		Film, 5~nm & 235 & $36.8590\pm0.0048$ & $-38.3518\pm0.5496$ & $-1.4928\pm0.5496$\\
		Film, 2.5-nm & 220  & $36.3827\pm0.0042$ &  $-19.6318\pm0.2460$ & $16.7509\pm0.2460$  \\
		Bulk & 220 &  $36.1575\pm0.0080$ & $-13.0510\pm0.2270$ & $23.1065\pm0.2270$\\
		Bulk & 225 &  $36.0145\pm0.0076$ & $-17.1152\pm0.2452$ & $18.8993\pm0.2452$ \\
		Bulk & 230 & $35.6952\pm0.0136$ & $-24.5543\pm0.3110$ & $11.1409\pm0.3110$\\
		Bulk & 235 &  $36.5120\pm0.0040$ & $-36.3790\pm0.3510$ & $0.1330\pm0.3510$\\
		\hline
	\end{tabular}
	\end{minipage}
\end{table*}

\subsection{Forward-flux Sampling Calculations\label{section:results:ffs}}

The 49 MD trajectories studied above only give us a phenomenological estimate of the likelihood of surface vs.~bulk crystallization. In order to obtain a more quantitative understanding, however, explicit calculations of nucleation rates are necessary. We thus use the forward flux sampling algorithm introduced above to compute nucleation rates in the very same films studied above (5-nm thick, $4\,096$ molecules). We perform these calculations at four  temperatures: 220, 225, 230 and 235~K. These are all significantly higher than the temperature of maximum crystallization rate, and as a result, spontaneous nucleation of ice in the supercooled liquid is very unlikely to occur at these temperatures. In order to quantify the effect of a flat interface on the nucleation rate, we perform the same rate calculation for a system that has no such interface, i.e.~the bulk system with equal number of molecules. These latter calculations are carried out at the same temperatures, and at a pressure of $p=1$~bar. 

Table~\ref{tab:flux} summarizes the technical specifications of the first stage of the FFS calculations aimed at computing fluxes. It is noteworthy that the computed fluxes are all of the same order of magnitude irrespective of temperature and the type of the system (bulk vs.~film). This is not surprising since the fluctuations that lead to these crossings are of thermal nature. By requiring the likelihood of crossing $\lambda_0$ to be between $10^{-3}$ and $10^{-2}$, we are implicitly fixing the number of trajectories that succeed in crossing $\lambda_0$. Therefore, the cumulative transition probabilities are good measures of (the order of magnitude) of nucleation rates. Fig.~\ref{fig:prob} depicts $\mathbb{P}(\lambda|\lambda_0)$ vs.~$\lambda$ for the bulk and the film calculations. Cumulative transition probabilities are consistently lower in the film than in the bulk at all temperatures considered in this work. Table~\ref{tab:rates} gives numerical values of the cumulative probabilities and rates alongside the error bars. Due to much smaller error bars in flux calculations, the uncertainty in computed nucleation rates mainly arises from the uncertainty in estimating the cumulative transition probabilities. Also note the eventual flatness of cumulative probability curves in Fig.~\ref{fig:prob}, which corresponds to the convergence of the FFS algorithm. Although the size of the critical nucleus at any given temperature and geometry can be determined from computing the commitor probabilities, one can obtain an upper bound by identifying the flat regions of the cumulative probability curves, since all the clusters in the flat region will be post-critical, otherwise they will have a nonzero probability of shrinking back to the liquid basin. 

Fig.~\ref{fig:rates} depicts the temperature dependence of the computed nucleation rates. The symbols correspond to the actual rates, while the curves are fitted according to  classical nucleation theory~\cite{GalliPCCP2011}. For the 5-nm films, the temperature dependence of  ice nucleation rates is similar to that of the rates in the bulk. This can be explained by the fact that the overwhelming majority of nucleation events that are sampled by the FFS algorithm involve crystalline nuclei that are partially located in the bulk. At large values of $\lambda$, these 'shared` clusters are more likely to grow in the bulk side than in the subsurface side. Consequently, the overall dynamics of crystallization is dominated by the underlying rate in the bulk, but is attenuated due to the unavailability of certain growth directions. This effect decreases the overall transition probabilities, as observed in Fig.~\ref{fig:prob}, but does not change the temperature dependence of rates in comparison to the bulk. This asymmetric growth into the bulk can be clearly seen in Fig.~\ref{fig:bulkfilm220} in which the average number of bulk and subsurface water molecules are depicted for the crystalline nuclei in configurations collected from rate calculations at 220~K. A similar behavior is observed at other temperatures, while the exact location of the crossover beyond which the subsurface portion of the largest cluster does not grow is different from temperature to temperature. We do not include the plots for other temperatures for conciseness reasons.

In order to factor out the impact of bulk-dominated asymmetric growth on nucleation, we construct a film that is 2.5~nm in thickness, and is therefore fully comprised of the subsurface region. We then use the FFS algorithm to compute the homogeneous nucleation rate in this 'ultra-thin` film. Due to high computational costs of  FFS calculations, we perform these calculations at one temperature only, namely at 220~K. Fig.~\ref{fig:thickness} depicts the cumulative transition probabilities for this ultra-thin film, as well as the 5-nm film and the bulk system at the same temperatures. 
 The fluxes and rates are also given in Tables~\ref{tab:flux} and~\ref{tab:rates}. Nucleation rates are about seven orders of magnitude smaller in the 2.5-nm ultra-thin film than in the bulk. This clearly shows the suppressive effect of the interface on ice nucleation, an effect that is partially masked in 5-nm films due to the dominance of asymmetric bulk-dominated crystallization.

\subsection{Free Energy Calculations}

\begin{figure}
	\begin{center}
		\includegraphics[width=.5\textwidth]{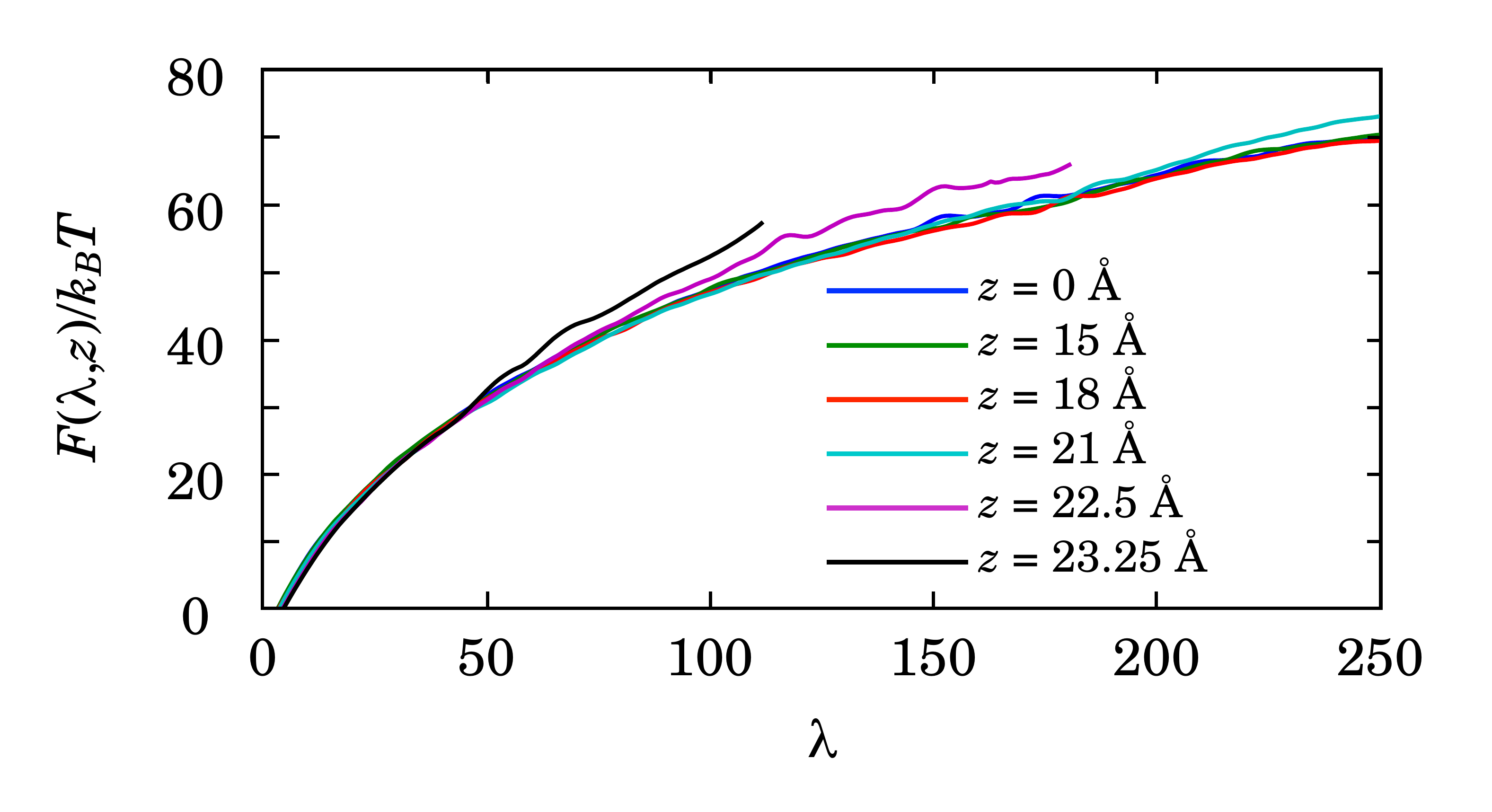}
		\caption{\label{fig:free_energy}Free energy of formation for crystalline nuclei of different sizes at different positions across the film. The distances are from the center of the film.}
	\end{center}
\end{figure}

In order to understand why a vapor-liquid interface suppresses ice nucleation at its vicinity, we use hybrid Monte Carlo and umbrella sampling to compute $F(\lambda,z)$, the free energy of formation for a crystalline nucleus of size $\lambda$ with its center of mass located at distance $z$ from the center of the 5-nm film. The temperature is set to 220~K. Due to the high computational cost of these calculations, we confine ourselves to clusters of 250 or fewer molecules as this range is sufficient for capturing the underlying physics of the nucleation process.  Fig.~\ref{fig:free_energy} depicts $F(\lambda,z)$ for different regions of the film. Each $F(\lambda,z)$ curve is obtained by averaging the two-dimensional free-energy surface in a slice that is centered at $z$ and is 0.3~\AA~thick. For small clusters, i.e.~the clusters with fewer than 50 water molecules, the free energy of formation is not sensitive to $z$.  
For larger clusters, however, the sensitivity starts to emerge in the subsurface region. For instance, a cluster of 100 water molecules at $z=23.25~\AA$ is around $5~k_BT$ less stable than a 100-molecule cluster located at the center of the film. This inferior surface stability penetrates deeper into the film as $\lambda$ increases. As can be seen in Fig,~\ref{fig:free_energy}, $F(\lambda,z)$ vs.~$\lambda$ is not sensitive to $z$ in the bulk region of the film determined in Fig.~\ref{fig:densitystress}. The inferior stability of large subsurface clusters are can partly explain the asymmetric growth observed in Fig.~\ref{fig:bulkfilm220}. It is necessary to mention that the calculations presented in this work overestimate the stability of the clusters that are in the subsurface region, since we do not prevent deformations of the vapor-liquid interface in our umbrella sampling calculations. Such deformations and ripples create clusters that have identical distances from the center of the film, but have different stabilities. This leads to an overestimation of the stability of surface clusters, and can also explain the numerical inaccuracies that can be observed in Fig.~\ref{fig:free_energy}.

\begin{figure*}
	\begin{center}
		\includegraphics[width=.95\textwidth]{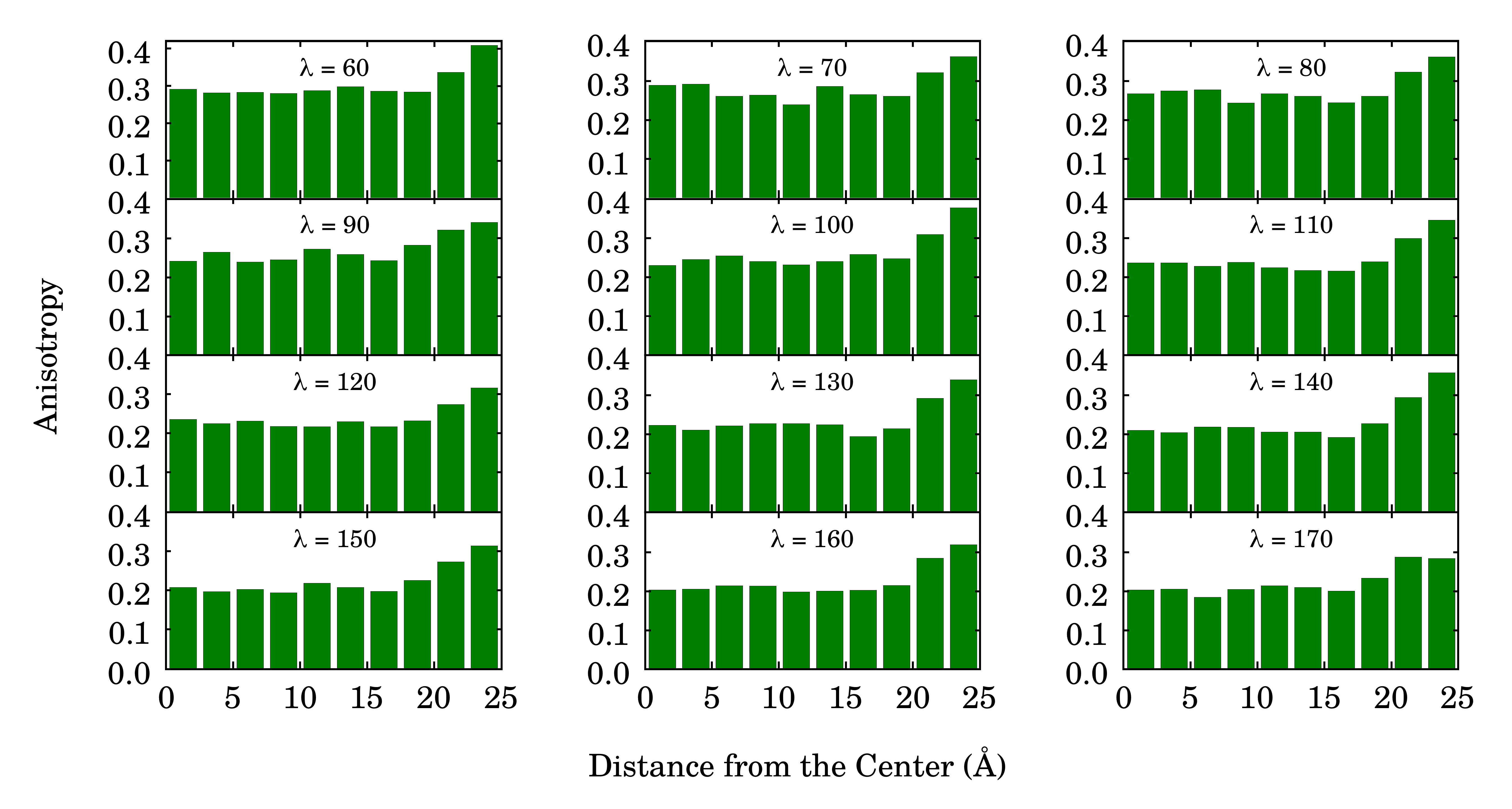}
		\caption{\label{fig:anise} Spatial distribution of the average anisotropies of the largest crystalline nuclei. Each histogram bin corresponds to a region of the film that is $2.5$~\AA~thick. Given distances are between the center of mass of the largest cluster and the center of the film. }
	\end{center}
\end{figure*}

\begin{figure}
	\centering
	\includegraphics[width=.5\textwidth]{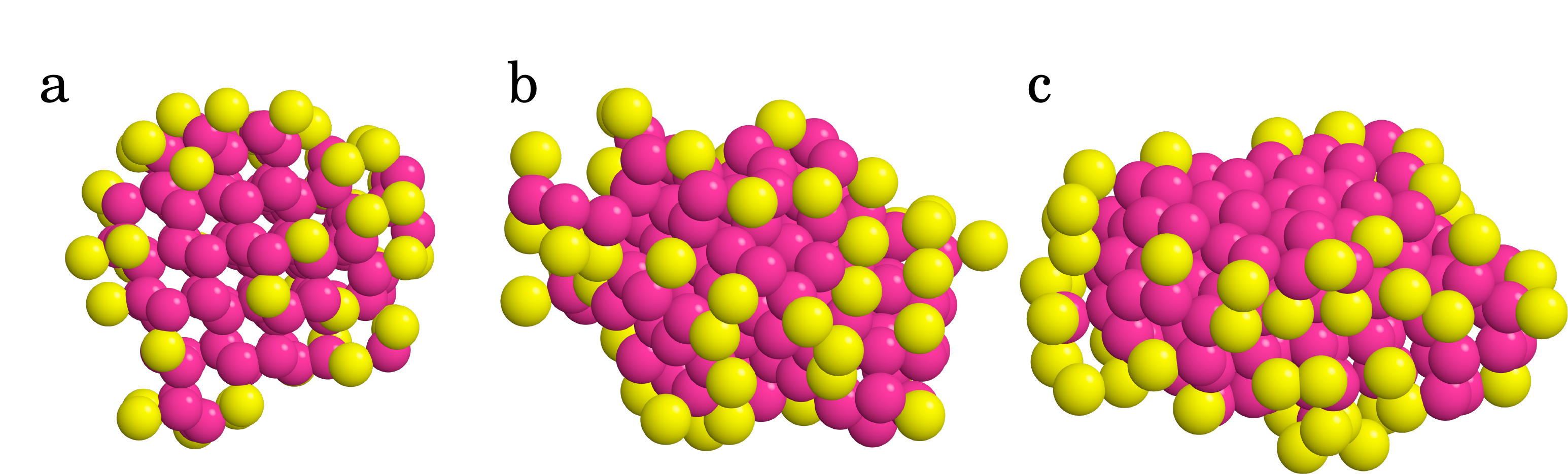}
	\caption{Crystalline clusters with different anisotropies: (a) a cluster of 174 water molecules with an anisotropy of 0.08 located at the center of the film, (b) a cluster of 174 water molecules with an anisotropy of 0.22 located at a distance of 12~\AA~from the center, and (c) a cluster of 175 water molecules with an anisotropy of 0.40 located at a distance of 20~\AA~from the center. The flat solid-vapor interface is visible at the top. In all these images, the pink molecules have a minimum of three solid-like neighbors.\label{fig:aniso_image}}
\end{figure}

Why are solid-like clusters less stable in the subsurface region? This question can be addressed both from a thermodynamic and a kinetic perspective. In general, what makes a pre-critical crystalline nucleus less stable is the free energy penalty associated with creating a solid-liquid interface. Different facets of a crystalline nucleus typically have different surface energies, but  this difference is usually not very large if all facets of the crystal are exposed to the same phase (e.g.~the liquid). Consequently, crystalline nuclei that are as spherical as possible are typically favored in the homogeneous nucleation of a crystal in the bulk liquid phase. This is not necessarily true when multiple amorphous phases are present in the system since different facets of the crystal might be exposed to vastly different environments, and can thus have vastly different energies of formation. The surface energies  needed for forming any of these facets will therefore be important in determining the geometry of crystalline nuclei, as well as the regime of volume- vs.~surface-dominated nucleation.  This is the theoretical basis of the theory proposed by Tabazadeh~\emph{et al.}~\cite{TabazadehPNAS2002} that was mentioned in Section~\ref{section:intro}. In this context, the reversible work needed for forming a solid-vapor interface in a two-phase liquid-vapor system is proportional to $\sigma_{sv}-\sigma_{lv}$, while the reversible work needed for creating a solid-liquid interface is proportional to $\sigma_{ls}$. 

For the mW system, these surface energies have been reported in the literature. Among them, $\sigma_{lv}$ is the easiest to compute, and has been calculated for a wide range of temperatures by Hudait~\emph{et al.}~\cite{MolineroJACS2014}. For 220~K, they report a value of $71~\text{mJ}\cdot\text{m}^{-2}$. Using the stress profiles~\cite{VegaTestAreaJCP2007} given in Fig.~\ref{fig:densitystress} to compute $\sigma_{lv}$, we are able to reproduce their results. $\sigma_{ls}$ and $\sigma_{sv}$ are however more difficult to compute.  Li~\emph{et al} utilized the classical nucleation theory to estimate $\sigma_{ls}$ and obtained a value of $31.01~\text{mJ}\cdot\text{m}^{-2}$~\cite{GalliPCCP2011}. Limmer and Chandler used a direct approach for computing $\sigma_{ls}$ in cylindrical nanopores~\cite{LimmerJCP2012}, and reported a value of $\approx30~\text{mJ}\cdot\text{m}^{-2}$ at $220$~K. In the case of $\sigma_{sv}$, the only available calculation is due to Hudait~\emph{et al.}~\cite{MolineroJACS2014} who performed conventional MD simulations to measure the contact angle of nanodroplets of mW water that are in contact with a sheet of ice  and use Young's equation to estimate $\sigma_{sv}$ from the computed contact angle, and the other surface tensions mentioned above. At the melting point, they observe a contact angle of $24^{\circ}$. By assuming that the contact angle is not a strong function of temperature, which is a reasonable assumption for most materials, one will get a solid-vapor surface tension of  $\sigma_{sv}\approx95~\text{mJ}\cdot\text{m}^{-2}$ at $220$~K. This will correspond to an energetic penalty of $30~\text{mJ}\cdot\text{m}^{-2}$ and $24~\text{mJ}\cdot\text{m}^{-2}$ for the formation of a solid-liquid and a solid-vapor surface respectively.  Due to the relatively close energetic penalties associated with the formation of a solid-liquid and a solid-vapor interface, one expects a strong correlation between the sphericity of a crystalline nucleus and its thermodynamic stability. For instance, if we assume that solid-liquid surface tension is not a function of $z$, and that the solid-vapor interface is flat, a hemispherical crystalline cluster of 150 water molecules that has a flat solid-vapor interface will be $\approx14~k_BT$ less stable than a spherical cluster of the same size completely immersed in the liquid. Of course, these assumptions are not accurate for the real system. However, this very simple calculation reveals how the larger surface-to-volume ratios of surface clusters tend to take over the slight energetic advantage of forming a vapor-liquid interface.

In order to test this hypothesis, we analyze over $600\,000$ configurations isolated from our umbrella sampling simulations and compute the anisotropy parameter, $\kappa$, from the gyration tensor of the largest crystalline nucleus of each configuration. If the eigenvalues of the gyration tensor are given by  $\gamma_1^2\ge\gamma_2^2\ge\gamma_3^2$, the anisotropy parameter $\kappa$, is defined as:
$$ \kappa^2 = \frac32\frac{\gamma_1^4+\gamma_2^4+\gamma_3^4}{(\gamma_1^2+\gamma_2^2+\gamma_3^2)^2}-\frac12	$$ 
For a collection of points in $\mathbb{R}^3$, $\kappa$ will vanish if those points are distributed uniformly inside a sphere. Therefore, larger values of $\kappa$ will correspond to distributions that are further away from such uniform distribution.  Fig.~\ref{fig:anise} depicts the spatial distribution of the average anisotropies of crystalline nuceli in different regions of the 5-nm film while Fig.~\ref{fig:aniso_image} shows representative clusters with different anisotropies. 
 The crystalline clusters that are close to the surface tend to be less spherical on average, which is consistent with our expectation. Visual inspection of these subsurface clusters reveals that they are predominantly hemispherical, with a flat solid-vapor interface (Fig.~\ref{fig:aniso_image}c).  

\begin{figure}
	\begin{center}
		\includegraphics[width=.5\textwidth]{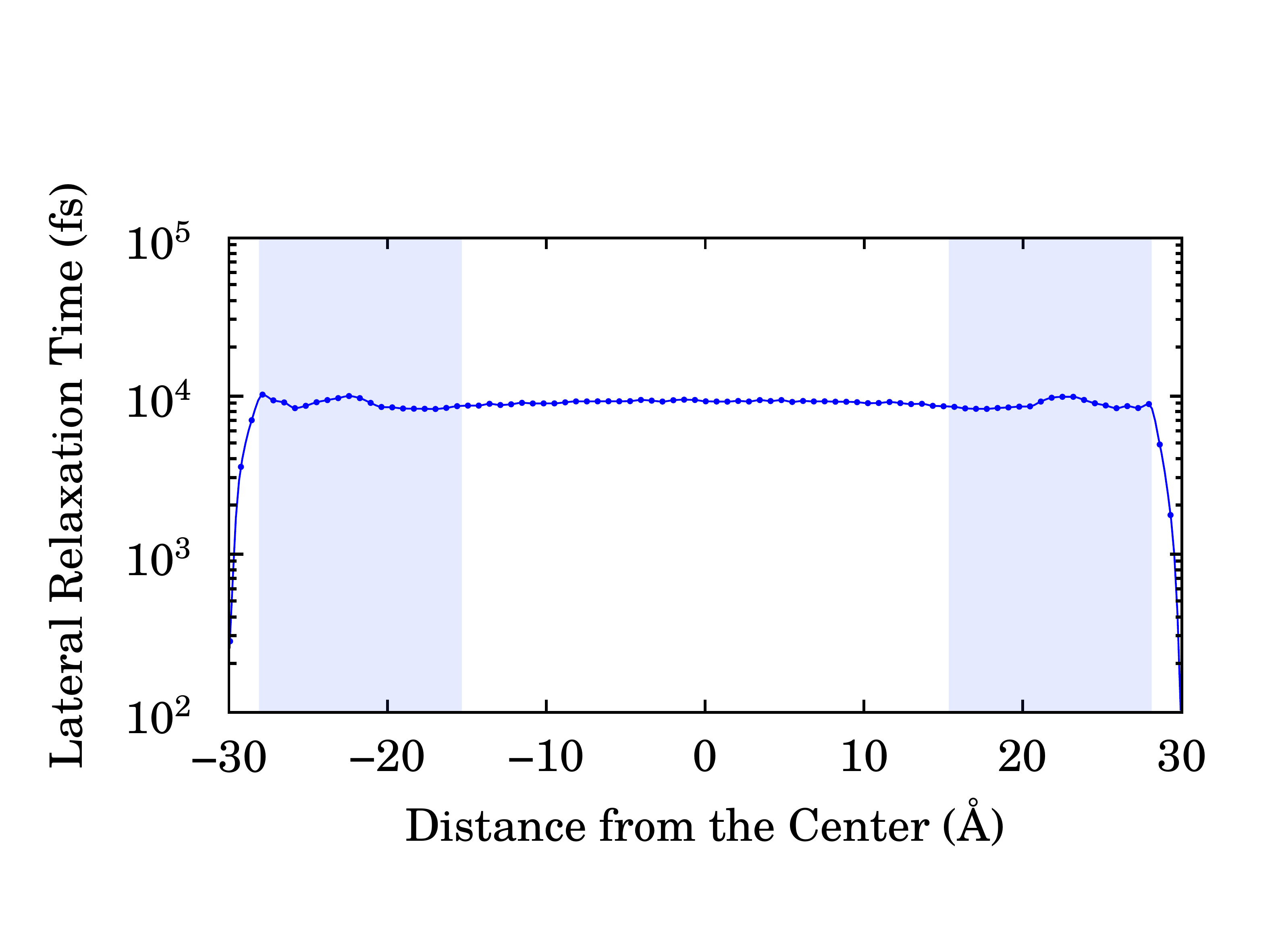}
		\caption{Lateral structural relaxation time vs. the distance from the center of the film for a thin film of $4\,096$ mW molecules at $220$~K. The shaded blue regions correspond to the subsurfaces of the vapor-liquid interfaces. Relaxation times are computed based on the decay of the self intermediate scattering function~\cite{HajiAkbariJCP2014}.\label{fig:rlx}}
	\end{center}
\end{figure}

Apart from this thermodynamic aspect that emanates from distinct geometries that form in different parts of the film, the kinetic behavior of the film might also be relevant to the observed suppression of crystallization in the mW system. Fig.~\ref{fig:rlx} shows the relaxation time profile across the 5-nm film computed from conventional MD simulations.  The technical details underlying this calculation are provided elsewhere~\cite{HajiAkbariJCP2014}. We observe that the relaxation time profile is fairly uniform across the film. Indeed, no subsurface region would have been detected if relaxation times had been used as the basis of the definition of the subsurface region. This behavior is distinct from what is observed in simple fluids, such as the Lannard-Jones system, where structural relaxation is significantly faster in the subsurface region than in the bulk~\cite{ShiJCP2011, HajiAkbariJCP2014}. The fact that dynamics is not faster in the subsurface region of the mW system deprives the subsurface region from its potential advantage over the bulk liquid, i.e.~its ability to harbor faster reconfiguration of molecules that are necessary for large density fluctuations.

\section{Conclusions}

In this work, we use molecular dynamics simulations and advanced sampling techniques and demonstrate that ice nucleation is suppressed in the vicinity of flat vapor-liquid interfaces for a coarse-grained monoatomic model of water, mW. The suppression of crystallization in the vicinity of curved vapor-liquid interfaces has been previously observed and has been attributed to the large Laplace pressure inside nano droplets of mW water~\cite{GalliNatComm2013}. Our explicit rate calculations reveal a decline in the nucleation rate of two to three orders of magnitude in films that are 5 nm thick, and a decline of seven orders of magnitude in films that are 2.5 nm thick. (This latter calculation has only been performed at 220~K.) Nucleation rates in the 5-nm films have a similar dependence on temperature as the rates in the bulk system, an observation that we attribute to the bulk-dominated asymmetric freezing in the film. We also use umbrella sampling simulations to estimate the  thermodynamic stability of crystalline nuclei of different sizes in different regions of the film, and conclude that the presence of the interface destabilizes  pre-critical crystalline nuclei in its vicinity. We explain this observation by analyzing the geometrical shapes of the clusters that form in different regions of the film, and observe that the clusters that are closer to the interface are more aspherical than the clusters that are in the bulk region. We also confirm that the pace of structural relaxation is uniform across the films, and no significant difference exists between the dynamics in the bulk and the dynamics in the subsurface region.

In Section~\ref{section:intro}, we discuss the theory of Tabazadeh~\emph{et al}~\cite{TabazadehPNAS2002}. In Section~\ref{section:results}, we use the reported surface tension values in the literature to compare the prediction of their theory to our observations. Although the inequality that they propose as a condition for surface-dominated crystallization is satisfied by the mW system, we observe a suppression-- and not a facilitation-- of freezing in the vicinity of liquid-vapor interfaces. This disagreement between the theory and simulation can be attributed to the tendency of the system to form hemispherical clusters at the interface due to the overall flatness of the original vapor-liquid interface. This increases the surface-to-volume ratio of the clusters that form at the interface in comparison to the clusters emerging in the bulk ($9/2r$ vs.~$3/r$). Therefore, the presumed energetic gain due to lower energetic penalties associated with the solid-vapor interface is offset by this increase in the surface-to-volume ratio. It thus appears prudent to revise this theory to account for the flatness of solid-vapor interfaces in systems where the energetic differences between competing solid-fluid interfaces are not very large.

In most materials, the solid phase is denser than the liquid. This is obviously not true for water, since the formation of a coherent tetrahedral network in ice creates void space in the crystal, making it less dense than the liquid. Therefore, the formation of ice can only proceed through density fluctuations that create locally dilute regions inside the liquid~\cite{ErrigngtonPRL2002}. This has led some to conjecture that the vapor-liquid interface will enhance crystallization in systems in which the liquid is denser than the crystal, since such density fluctuations  would tend to occur with greater ease in the vicinity of a vapor-liquid interface. Earlier computer simulations of silicon~\cite{LiNatMater2009}, another tetrahedral fluid with a liquid denser than its crystal, revealed that crystallization is indeed enhanced in the subsurface region. Our calculations clearly demonstrate that this conjecture is not true, and the effect of a vapor-liquid interface on crystallization appears to be too complex  to be rationalized solely on the basis of parameters such as the density difference between the liquid and the solid.

One of the most important characteristics of the mW model that makes it very popular in computational studies of water is its lack of electrostatic interactions. This not only reduces the amount of computer time needed for integrating Newton's equations of motion, but also accelerates the intrinsic dynamics of the mW system in comparison to molecular-- i.e. multi-site-- models of water because the pace of structural relaxation in molecular models is hampered by the slowness of rotational rearrangements of molecules that are necessary for the rearrangement of the hydrogen bond network. As rewarding as it might be for most applications, this feature is likely to become a shortcoming in studying confined systems, as it will mask charge imbalances that are likely to develop at interfacial regions. Indeed, the earlier computational studies of Jungwirth~\emph{et al}.~\cite{JungwithJPCB2006, JungwirthJPhysChemC2010} reveal the existence of these charge imbalances at vapor-liquid interfaces and their potential role in promoting crystallization in free-standing thin films of molecular water.  Although the water model used by Jungwirth~\emph{et al.} is not among the most accurate ones, it demonstrates the possibility that electrostatics might play an important role in crystallization at interfaces. What we are able to establish in this work is the fact that local tetrahedrality in a water model does not necessarily lead to the enhancement of crystallization in the vicinity of the vapor-liquid interface. Whether the presence of electrostatics will lead to the enhancement of crystallization in the subsurface region can only be addressed by repeating the current study for a good molecular model of water such as TIP4P/2005~\cite{VegaTIP4P2005} or TIP4P/Ice~\cite{VegaTIP4PiceJCP2005}. As mentioned in Section~\ref{section:intro}, the problem of computing nucleation rates for molecular models of water is, however, very challenging and has not  been solved, even for homogeneous nucleation of ice in  bulk supercooled water. Until this long-standing challenge is overcome, studying the role of electrostatics in enhancing or suppressing ice nucleation in the vicinity of interfaces, using realistic, multi-site models of water, will remain beyond reach. 

\acknowledgments
P.G.D. and A.H.A. gratefully acknowledge the support of the National Science
Foundation (Grant No.~CHE-1213343) and of the Carbon Mitigation Initiative at Princeton University (CMI). S.S. and R.S.D. gratefully acknowledge the support
of NSF (Grant No.~ACI-1212680). These calculations were partly performed on the Terascale Infrastructure for Groundbreaking Research in Engineering and Science (TIGRESS) at Princeton University.  This work used the Extreme Science and Engineering Discovery Environment (XSEDE), which is supported by National Science Foundation grant number ACI-1053575. We gratefully acknowledge R. Allen and V. Molinero for useful discussions.

\bibliographystyle{rsc}
\bibliography{References2}

\end{document}